\ifdefined\BUILDRTGBOOK
\else
\documentclass{rtg}
\fi

\rtgTitle{Real-time rendering of complex fractals}
\rtgAuthor{Vinícius da Silva,$^1$ Tiago Novello,$^2$ Hélio Lopes,$^1$ and Luiz Velho$^2$\\
$^1$PUC-Rio\\
$^2$IMPA}

\rtgBegin

\rtgAbstract{
This chapter describes how to use intersection and closest-hit shaders to implement real-time visualizations of complex fractals using distance functions. The \emph{Mandelbulb} and \emph{Julia Sets} are used as examples.
} \rtgBibStart

\section{Overview}
\label{sec:overview}

Complex dynamics fractals have interesting patterns which can be used to create effects or mood in 3D scenes. In 2014, Walt Disney Animation Studios used the interior of the Mandelbulb fractal in the computer-animated movie \textit{Big Hero 6}~\cite{hutchins2015big} to design the inner of a wormhole. Marvel Studios also used Mandelbulbs to produce the magical mystery tour scene in the Doctor Strange movie~\cite{drstrange20}.

\subsection{Julia sets}
Examples of \textit{Julia sets} arise from the exploration of the convergence of the system given by the iterations of the quadratic function $f(z)=z^2+c$. Specifically, a (filled-in) Julia set consists of the set of points $z_0$ in the complexes/quaternions, where the sequence $f^n(z_0)$ has a finite limit. Changing the constant $c$ produces different Julia sets. 

Using the complex plane as the domain of the quadratic function $f$, results in the traditional images of the 2D Julia sets. Norton~\cite{norton1989julia} extended this class of fractals to 4D considering that the quaternions are the domain of $f$. The resulting 4D Julia set can be seen in the 3D space by taking 3D slices of the quaternions.

To visualize a 2D Julia Set, we determine whether a point on the complex plane diverges. Therefore, it suffices to compute the sequence $f^n(z_0)$ and see how quickly its magnitude increases. This test can be applied on points corresponding to pixels in an image, resulting in an illustration of a 2D Julia set.

The above approach is not efficient to render a 3D Julia set. However, as we are interested in looking at the fractal "surface", ray tracing seems to be an appropriate technique. Ray tracing fractals dates back to the work of Hart et al.~\cite{hart1989ray} which uses a distance estimator (described in~\cite{norton1989julia}) to speed up the ray tracing process using \textit{unbounding spheres}. Recently, Quilez~\cite{Quilez20} presented real-time visualizations of the 3D Julia sets using pixel shaders, applying techniques similar to those defined in~\cite{hart1989ray}.

Inspired by the work of Quilez, we use DXR shaders to produce visualizations of the 3D Julia set. We generate visualizations based on cutting the 3D Julia set using a plane. See Figure~\ref{fig:cutJulia}. 

\begin{figure}[!ht]
    \centering
    \includegraphics[width=0.8\textwidth]{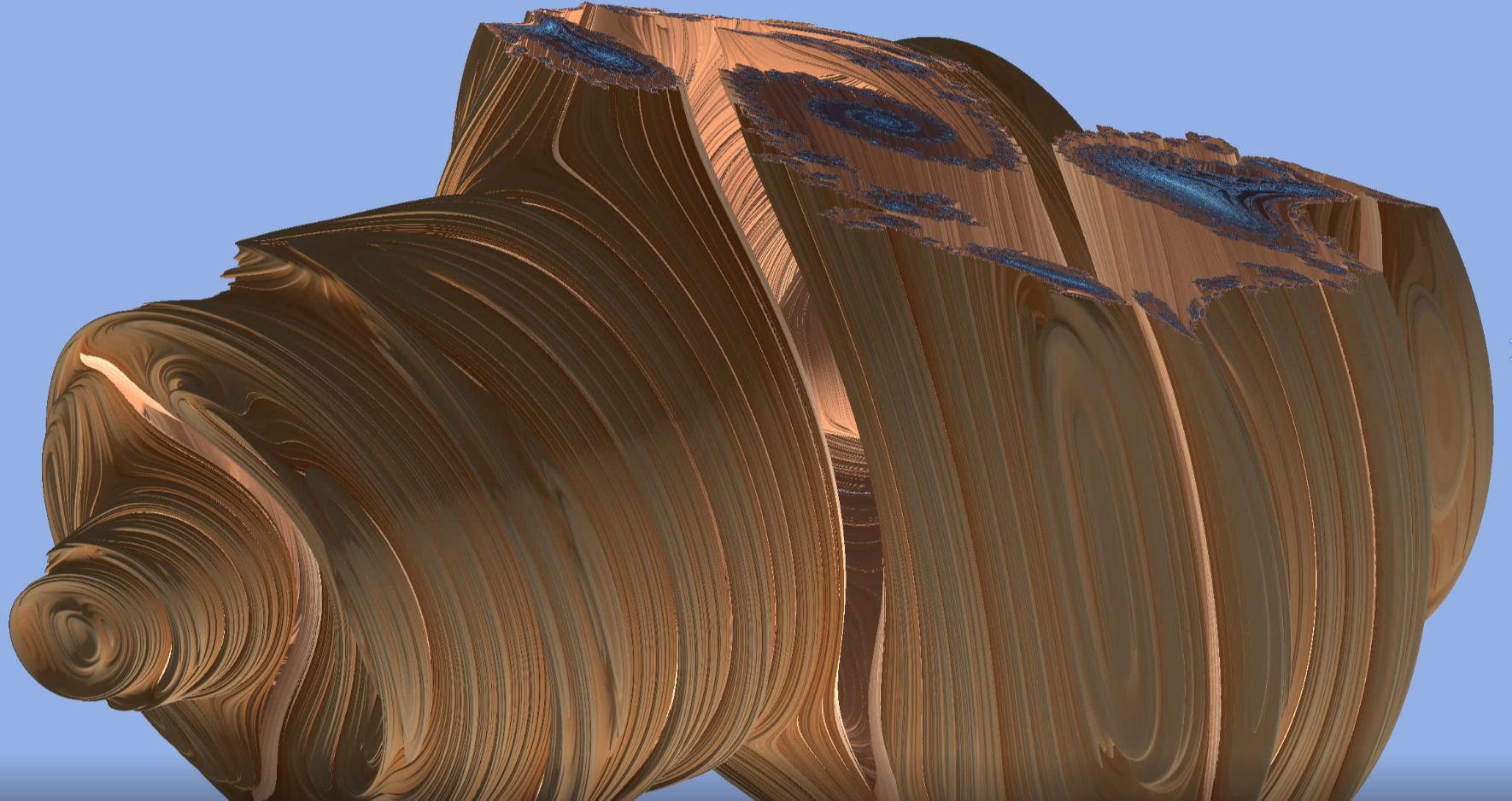}
    \includegraphics[width=0.8\textwidth]{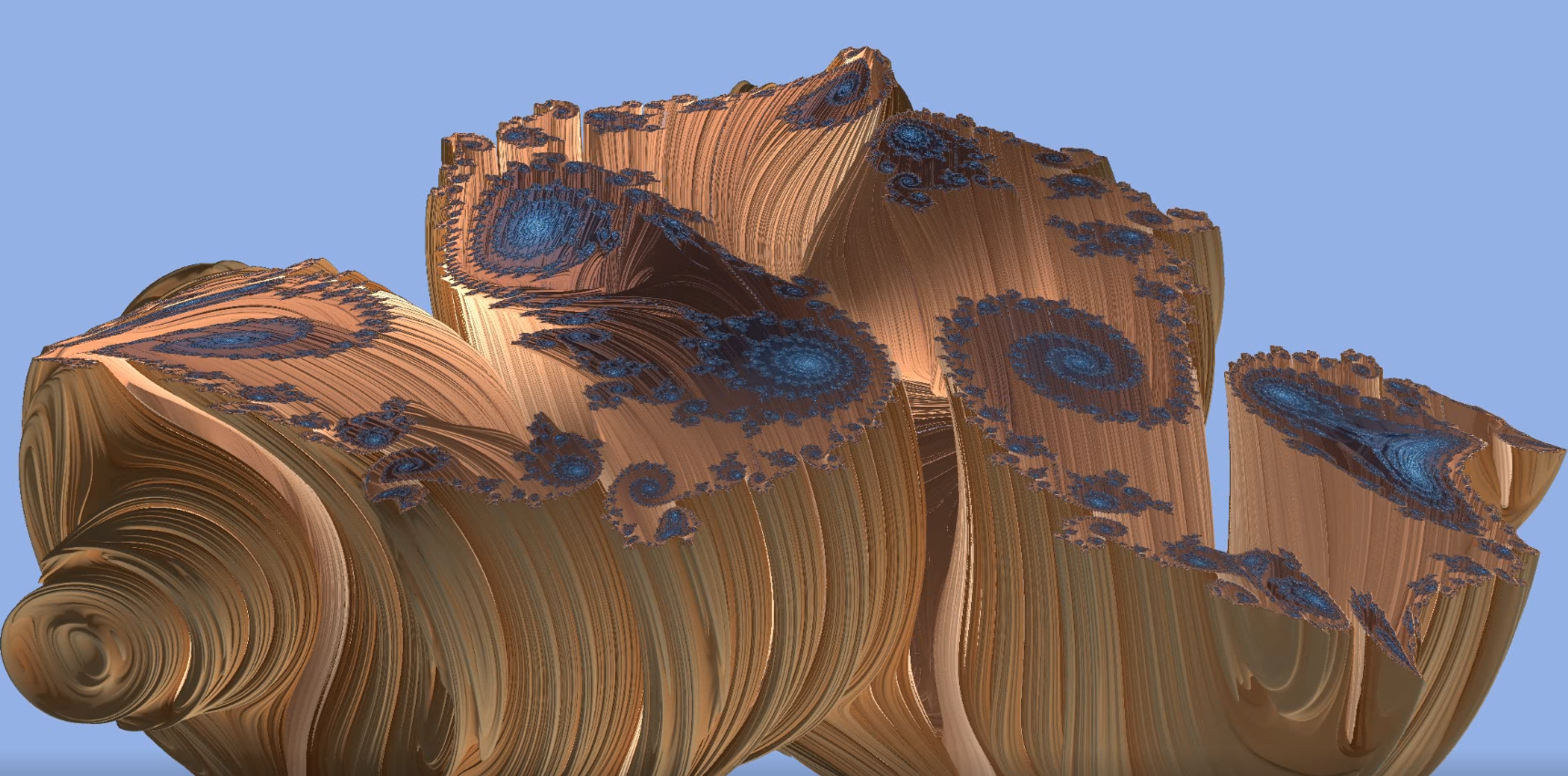}
    \caption{Two Illustrations of the 3D Julia set cut by two different planes. The images restricted to 2D slides are similar to the 2D Julia sets. Rendered using the shaders described in this chapter.}
    \label{fig:cutJulia}
\end{figure}

\subsection{Mandelbulb}\label{s-mandelbulb}
The \textit{Mandelbulb} is a 3D fractal, constructed recently by Daniel White~\cite{white2017mandelbulb} and Paul Nylander~\cite{nylander2017mandelbulb}. It is a commonly used representation of a 3-dimensional Mandelbrot set. A canonical representation does not exist because there is no 3-dimensional analogue of the 2-dimensional space of complex numbers or quaternions, different from the Julia Sets. We hope the discussion and images presented in this chapter motivate a deep instigation of the Mandelbulb by the mathematical community. 

White and Nylander considered the geometrical properties of the complex numbers (multiplication is related to rotation and addition becomes a movement in a particular direction) to define a kind of "product" of elements in the 3D space. Using this "multiplication" (in 3D) in the polynomial formula $f(z)=z^n+c$ leads us to the Mandelbulb. Specifically, the Mandelbulb is defined as the set of points $c$ in $\mathbf{R}^3$ such that the sequence $f^n(0)$ (the orbit of $0$) is bounded, i.e. the sequence does not diverge. This definition is very similar to the well-known \textit{Mandelbrot fractal} in the complex case.

The White and Nylander's formula for the ``$n$th power'' of a point $p=(x,y,z)$ is 
\begin{equation}\label{eq-white_formula}
    p^n:=r^n( \sin(n\theta)\cos(n\phi), \sin(n\theta)\sin(n\phi) , \cos(n\theta)).
\end{equation}
Where $r=|p|$ is the norm of $p$ and $\theta = \arctan(y/x)$ and $\phi=|(x,y)|/z$ are the spherical coordinates of $p/|p|$. This ``product'' was defined using an extension of the $n$th power of complex numbers. We give the motivation of this formula. Let $z=r(\cos(\theta)+i\sin(\theta))=r\cdot e^{i\theta}$ be a complex number represented using Euler's formula. Therefore, the $n$th power of $z$ is given by $z^n=r^n\cdot e^{i(n\theta)} = r^n(\cos(n\theta)+i\sin(n\theta))$.

Figure~\ref{fig:mandelbulb} illustrates the Mandelbulb represented by the function $f(p)=p^8+c$ using Equation~\ref{eq-white_formula}. 

\begin{figure}[!ht]
    \centering
    \includegraphics[width=0.4\textwidth]{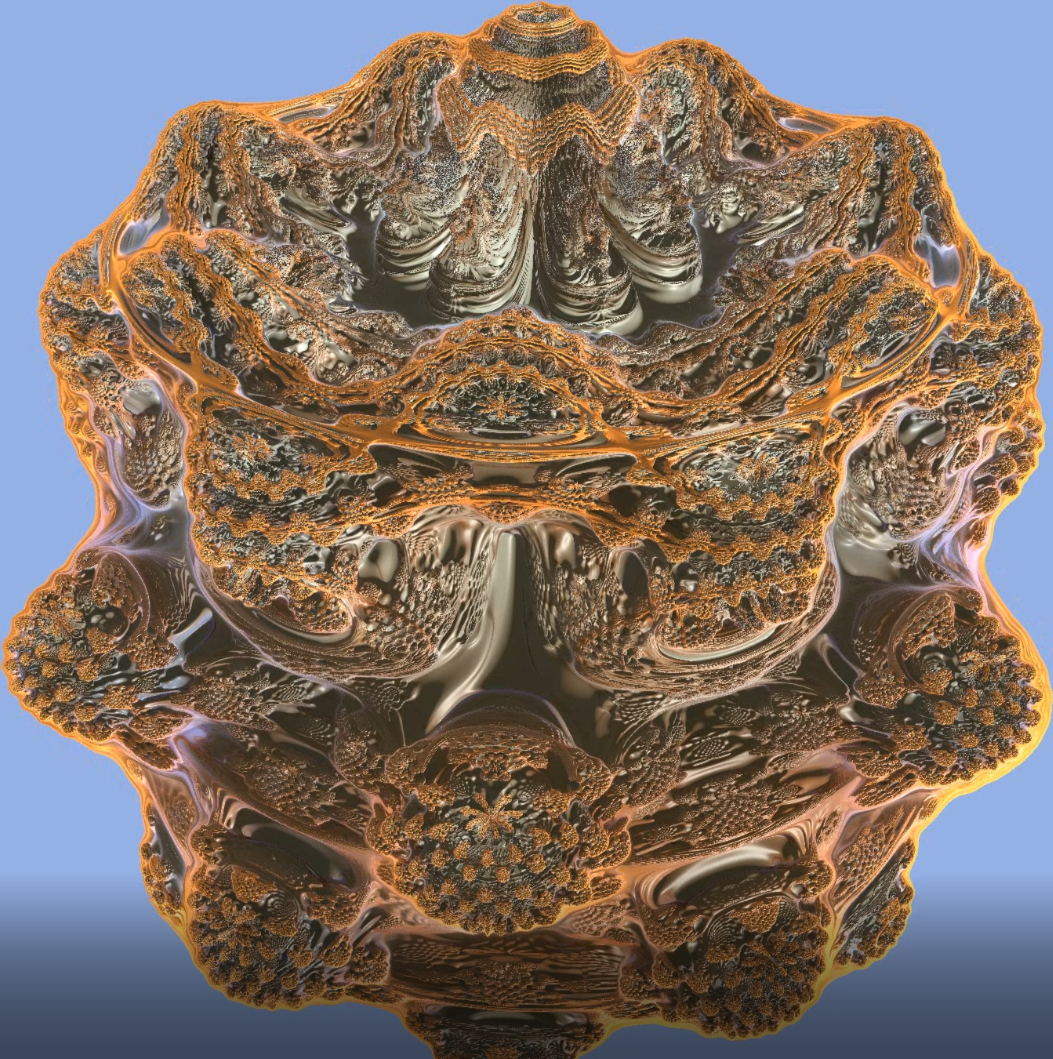}
    \includegraphics[width=0.45\textwidth]{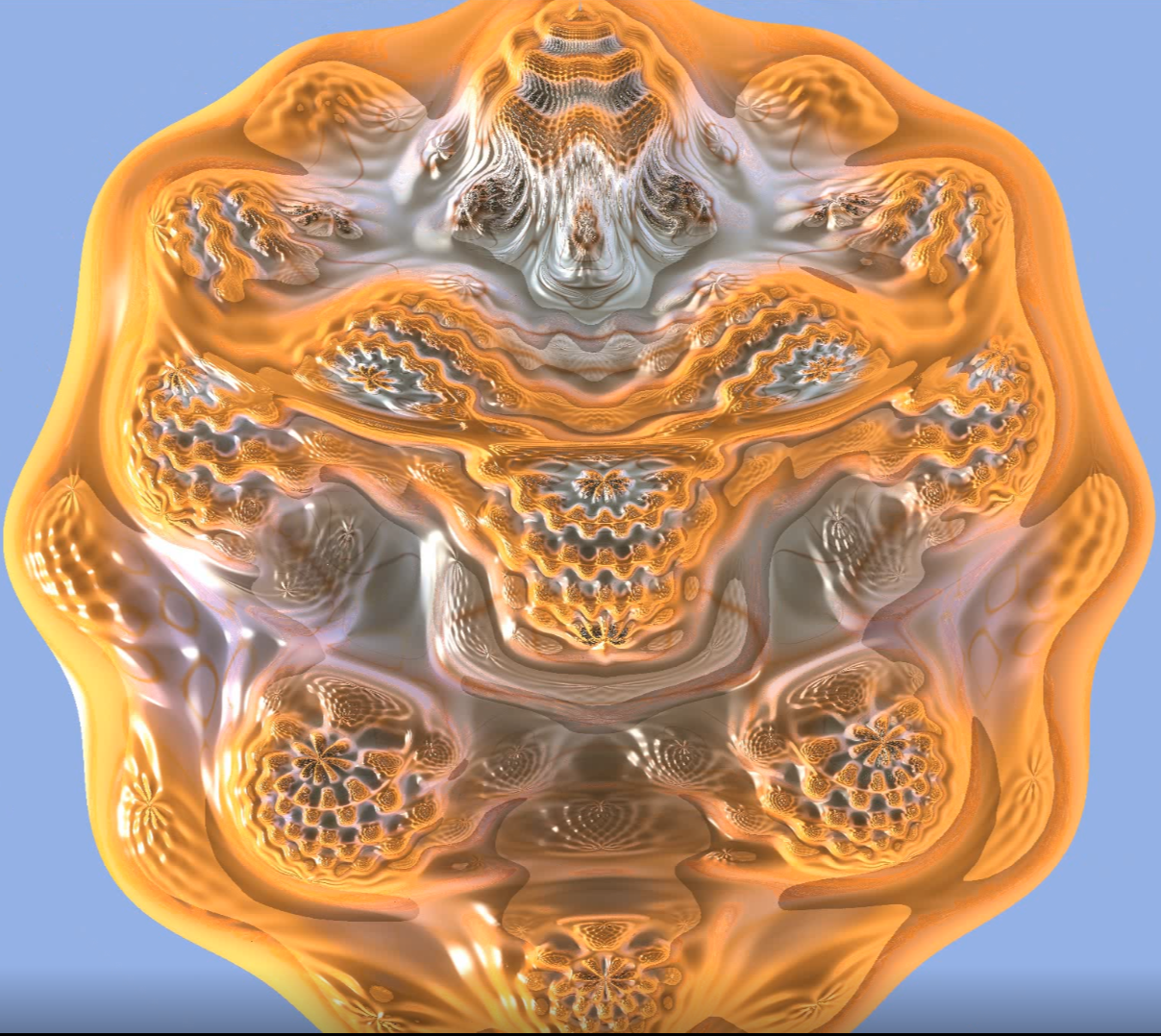}
    \caption{Two frames of an animated Mandelbulb rendered using the shaders described in this work, considering $n=8$ in $f(z)=z^n+c$. The animation is created varying the number of iterations used in the ray marching over time.}
    \label{fig:mandelbulb}
\end{figure}

\section{Distance functions}
In this section, we present an approximation of the distance function of the Julia set surface.
We follow the definition and notations of Quilez~\cite{Quilez2004}, presenting a non-rigorous but intuitive discussion about it.

Let $f(z)=z^p+c$ be a polynomial function of degree $p$. The domain of $f$ could be the complexes or the quaternions. Our objective is to define a distance from the set of points $z_0$ in the complexes/quaternions, where the sequence $f^n(z_0)$ has a limit. The expression $f^n$ means that $f$ is composed $n$ times, i.e. $f^n(z)=(f^{n-1}(z))^p+c$. The \textit{Boettcher map} $\phi_c(z)=\lim_{n\to \infty} (f^n(z))^{p^{-n}}$ is used to derive the distance function. When considering the complexes, this map is a deformation of the complex plane.

We provide an informal explanation of the importance of the Boettcher map. Let $z$ be an element such that the sequence $f^n(z)$ diverges, i.e. $\lim_{n\to \infty}|f^n(z)|\to \infty$. Then, for a $n$ big enough, $f^n(z)$ is far away from the Julia set. In this case, the term $f^n(z)$ will dominate over $c$. Therefore, we can forget about $c$ in the expression $f(z)=z^p+c$ to obtain $f^{n+1}(z)\approx (f^{n}(z))^p$. 
In this case, the expression of $f$ turns out to be $f_0(z)=z^p$, so we can undo the interactions by considering $(f_0^{n}(z))^{p^{-n}}=z$. Therefore, $\phi_0(z)=z$ when $f^n(z)$ diverges. 


According to the aforementioned property of $\phi_0(z)$, the map approximates to the identity $\lim_{z\to\infty}\phi_c(z)=z$ as we move away from the Julia Set. Figure~\ref{fig:julia2d} illustrates this situation for $f(z)=z^2+c$. Note that the regions distant from the Julia set (red square) receive less deformation, while the regions near it (green square) deform more.

\begin{figure}[!ht]
    \centering
    \includegraphics[width=0.8\textwidth]{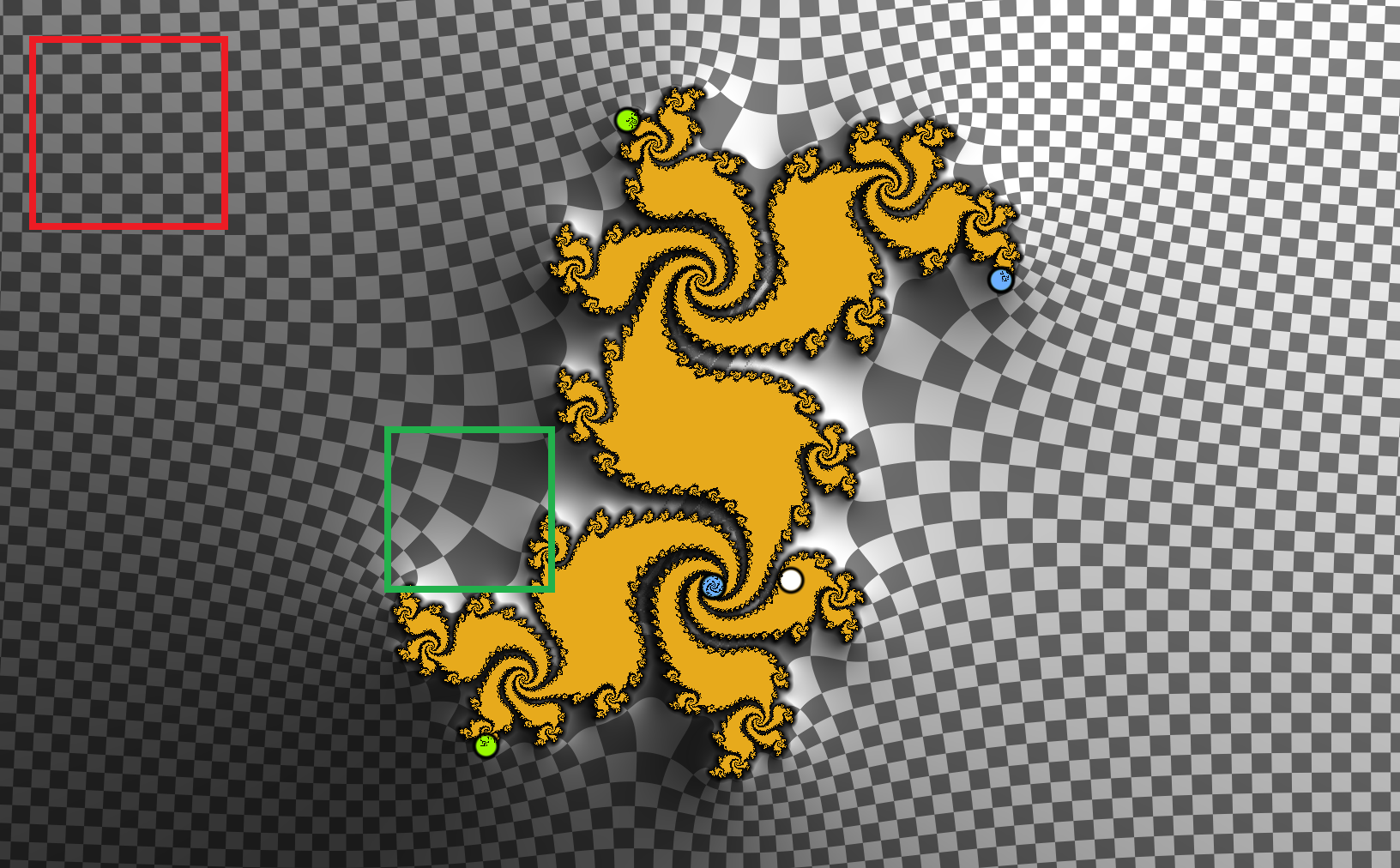}
    \caption{Boettcher map of the Julia Set. The red square shows an area with small deformations, where $\phi_c(z) \approx z$. The green square shows an area near the Julia Set, where deformations are stronger. The colored points can be mostly ignored for our purposes (the white point is $c$, and the other ones are fixed points of the dynamics). Image generated using Inigo Quilez's implementation in Shader Toy~\cite{Quilez13}.}
    \label{fig:julia2d}
\end{figure}

Based on the Boettcher map, we now search for a function that represents distance, which takes complexes/quaternions and returns a positive number. For this, we define $G(z)=\log|\phi_c(z)|=\lim_{n\to \infty}\frac{\log|f^n(z)|}{p^n}$. It is easy to see that this function is smooth and is 0 at the Julia set because $f^n(z)$ converges and $p^n$ grows exponentially. In addition, as we move away from the Julia set, $G(z)$ tends to $\log|z|$ since $\lim_{z\to \infty}\phi_c(z)=z$. Finally, because of its continuity, the function $G$ gets close to zero as we get close to the set. Even though $G$ is not necessarily a distance function (because it tends to be $\log|z|$ far away from the Julia set), it behaves like one.

To make the distance function we use an approximation based on the first order Taylor expansion of $G$. As a result, we obtain an upper bound estimation $d(z)=\frac{|G(z)|}{|\nabla G(z)|}$, which uses $G$ and its gradient $\nabla G$. 
It is not mandatory, but we explain this estimator for completeness.

Let $z$ and $\epsilon$ be points in the complexes/quaternions such that $z+\epsilon$ is the closest point in the Julia set to $z$, i.e. $|\epsilon|$ is the desired distance. Then, we have that $G(z+\epsilon)=0$, so by the first order Taylor expansion of $G$, we obtain $0=G(z)+\langle\nabla G(z), \epsilon\rangle +O(|\epsilon|^2)$. Let's assume that the approximation is ``precise'', i.e. $0=|G(z)+\langle\nabla G(z), \epsilon\rangle|$. Then, using the following inequalities $|G(z)+\langle\nabla G(z), \epsilon\rangle|\geq |G(z)|-|\langle\nabla G(z), \epsilon\rangle|$ and $|\langle\nabla G(z), \epsilon\rangle|\leq |\nabla G(z)| |\epsilon|$, we get an upper bound to the distance $|\epsilon|\geq\frac{|G(z)|}{|\nabla G(z)|}$. We derive the first inequality from the \textit{triangle inequality}, and the second comes from the \textit{Cauchy--Schwarz inequality}.

Using the definition $G(z)=\lim_{n\to\infty}\frac{\log|f^n(z)|}{p^n}$ and its derivatives $\nabla G(z)=\frac{|(f^n)'(z)|}{p^n|f^n(z)|}$, the distance function $d(z)=\frac{|G(z)|}{|\nabla G(z)|}$ can be expressed as 
\begin{equation}\label{eq-distance}
d(z)= \lim_{n\to\infty}\frac{|f^n(z)|\log|f^n(z)|}{|(f^n)'(z)|}.    
\end{equation}

From Equation~\ref{eq-distance}, we have to compute both $f^n$ and its derivative during the iteration loop.
We compute the sequences using
\begin{eqnarray}
f^{n+1}&=&(f^n)^p+c\label{eq-iteration}\\
(f^{n+1})'&=&p\cdot 
(f^n)^{p-1}\cdot(f^n)'\label{eq-derived_iteration}
\end{eqnarray}

Equation~\ref{eq-iteration} is the recursion corresponding of iterating $f$, and its derivative is presented in Equation~\ref{eq-derived_iteration} with the initial condition $(f^0)'=1$, since $f'(z)=p\cdot z^{p-1}$.
Iterating these equations results in an algorithm (see Listing~\ref{code:julia_map}) to compute the distance function in Equation~\ref{eq-distance}.

\vspace{0.2cm}

To do the shading of the point $z$ of the Julia set, we need an approximation of its normal vector. As the function $d(z)=\frac{G(z)}{|\nabla G(z)|}$ is an approximation of the distance function from the Julia set, its gradient provides the desired normal vector. 
We could compute the gradient analytically using the formula $\nabla d=\frac{\nabla G|\nabla G|-G\nabla|\nabla G|}{|\nabla G|^2}$, which comes from the quotient rule of the gradient. When restricted to the Julia Set, the expression turns out to be $\nabla d=\frac{\nabla G}{|\nabla G|}$, since the function $G$ is 0 at the set. 
Instead, we consider a simple numerical approach for convenience.

We use an efficient procedure to numerically compute the gradient of a function $f:\mathbf{R}^3\to \mathbf{R}$ to compute $\nabla d$.
Let $\nu_0=(1,-1,-1)$, $\nu_1=(-1,-1,1)$, $\nu_2=(-1,1,-1)$, and $\nu_3=(1,1,1)$ be the vertices of a tetrahedron.
The gradient $\nabla f(p)$ can be approximated by the formula

\begin{equation}\label{eq-gradient}
    \nabla f(p) \approx \displaystyle \frac{1}{4h}\sum_i \nu_i f(p+h\nu_i).
\end{equation}

To derive that expression, we define $m:=\sum_i \nu_i f(p+h\nu_i)$, and rewrite it using that $\sum_i v_i=(0,0,0)$ to obtain $m=\sum_i \nu_i (f(p+h\nu_i)-f(p))$. Using the approximation $\frac{\partial}{\partial \nu_i}f(p)\approx \frac{f(p+h\nu_i)-f(p)}{h}$ for the derivative of $f$ in the direction $\nu_i$, we get $m\approx \sum_i \nu_i\cdot h \cdot \frac{\partial}{\partial \nu_i}f(p)$. From calculus, we have $\frac{\partial}{\partial \nu_i}f(p)=\langle \nu_i, \nabla f(p) \rangle$, therefore $m\approx \sum_i \nu_i\cdot h \cdot \langle \nu_i, \nabla f(p) \rangle$. We look at the component $x$ of $m$; the computations are analogous for $m_y$ and $m_z$. 
$$\frac{m_x}{h} \approx \sum_i (\nu_i)_x \cdot \langle \nu_i, \nabla f(p) \rangle = \langle \sum_i (\nu_i)_x\nu_i , \nabla f(p) \rangle=\langle (4,0,0), \nabla f(p)\rangle$$
We used the linearity of the dot product in the second equality. It is easy to verify that $\sum_i (\nu_i)_x\nu_i=(4,0,0)$ which explains the last equality. As a result, we have $\frac{m}{4h}\approx\nabla f(p)$ as stated in Equation~\ref{eq-gradient}.

\section{Implementation}
\label{sec:implementation}

We assume several procedural objects in the scene, each one with a matrix transforming world space into AABB's local space. The shaders were implemented in HLSL, using the Proceduray engine~\cite{dasilva2020proceduray}. The host setup is beyond the scope of this chapter, but we refer to that reference as an in-depth guide to do so.

\subsection{Julia Sets}\label{s-julia_implementation}

Listing~\ref{code:intersection} shows the code for the intersection shader of the Julia Set. 
The function \verb|GetRayInLocalSpace()|, in line 4, computes the origin and direction in the local coordinates of the underlying AABB.
The corresponding ray is passed to \verb|JuliaDistance()| (Listing~\ref{code:julia_distance}), in line 8, which determines the ray parameter corresponding to the intersection between the ray and the Julia set and the normal at the hit point. \verb|thit| is a \verb|float2| because it also contains the number of iterations in the distance function. Those values are used latter in the closest-hit shader (Listing~\ref{code:closest}) to compute the shading of the object.

\pagebreak

\begin{lstlisting}[caption={Intersection shader.}, label={code:intersection}, language=rtgCode,
	morekeywords={[1]Intersection_Julia, GetRayInLocalSpace, JuliaDistance, RayTCurrent, mul, WorldToObject3x4, ReportHit},
	morekeywords={[2]shader, ProceduralPrimitiveAttributes, Ray, float4, float3x3}]
[shader("intersection")]
void Intersection_Julia()
{
    Ray ray = GetRayInLocalSpace();
    float2 thit;
    ProceduralPrimitiveAttributes attr;
    float3 pos;
    bool test = JuliaDistance(ray.origin, ray.dir, attr.normal, thit);
    if (test && thit.x < RayTCurrent())
    {
        attr.normal = mul(attr.normal, (float3x3) WorldToObject3x4());
        attr.color = float4(thit, 0.f, 0.f);
        ReportHit(thit.x, /*hitKind*/ 0, attr);
    }
}
\end{lstlisting}

\verb|JuliaDistance()| (Listing~\ref{code:julia_distance}) returns a boolean indicating if there is an intersection and outputs \verb|resT| and \verb|normal|. It calls functions for finding the distance using raycasting (Listing~\ref{code:julia_raycast}) in line 5 and to calculate the normals at the intersection point (Listing~\ref{code:julia_normals}) in line 10.

\begin{lstlisting}[caption={JuliaDistance.}, label={code:julia_distance}, language=rtgCode,
	morekeywords={[1]in, inout},
	morekeywords={[2]JuliaDistance, raycast, calcNormal}]
bool JuliaDistance(in float3 ro, in float3 rd, inout float3 normal,
    inout float2 resT)
{
    resT = 1e20;
    float2 tn = raycast(ro, rd);
    bool cond = (tn.x >= 0.0);
    if (cond)
    {
        float3 pos = (ro + (tn.x * rd));
        normal = calcNormal(pos);
        resT = tn;
    }
    return cond;
}
\end{lstlisting}

\verb|raycast()| (Listing~\ref{code:julia_raycast}) computes an approximation of the first intersection. The algorithm delimiters an intersection search interval at lines 1 and 2. In line 5, it updates that interval based on a bounding sphere and two clipping planes (to cut the Julia Set, as in Figure~\ref{fig:cutJulia}).
The raycast loop (lines 11-19) uses function \verb|dist()| (Listing~\ref{code:julia_map}), in line 13, to calculate the distance. 

\pagebreak

\begin{lstlisting}[caption={Julia Set ray cast.}, label={code:julia_raycast}, language=rtgCode,
	morekeywords={[1]in},
	morekeywords={[2]raycast, checkBoundaries, dist}]
float2 raycast(in float3 ro, in float3 rd)
{
    float tmin = kPrecis;
    float tmax = 7000.f;
    if (!checkBoundaries(ro, rd, tmin, tmax))
        return float2(-2.0, 0.0);
    float2 res = { -1.0, -1.0 };
    float t = tmin;
    float lt = { 0.0 };
    float lh = { 0.0 };
    for (int i = 0; i < 1024; i++)
    {
        res = dist(ro + (rd * t));
        if (res.x < kPrecis) break;
        lt = t;
        lh = res.x;
        t += min(res.x, 0.2);
        if (t > tmax) break;
    }
    res.x = (t < tmax) ? t : -1.0f;
    return res;
}
\end{lstlisting}

\verb|dist()| (Listing~\ref{code:julia_map}) implements Equations~\ref{eq-iteration} and \ref{eq-derived_iteration}, used to approximate the distance function given in Equation~\ref{eq-distance}.
The variable \verb|z|, defined in line 3, is a \verb|float4| representing a quaternion that is the initial condition of the recursion in Equation~\ref{eq-iteration}. Line 4 defines the initial condition of the recursion in Equation~\ref{eq-derived_iteration}.
The loop (lines 7-14) does the iterations of the system (Equations \ref{eq-iteration} and \ref{eq-derived_iteration}). There is a break, in line 12, to stop the loop when the sequence given by the iterations of \verb|z| (line 10) diverges.
Finally, line 15 computes an approximation of the distance function using Equation~\ref{eq-distance}.

\begin{lstlisting}[caption={Julia Set distance.}, label={code:julia_map}, language=rtgCode,
	morekeywords={[1]in},
	morekeywords={[2]dist, qLength2, qCube, qLength2}]
float2 dist(in float3 p)
{
    float4 z = float4(p, 0.0);
    float dz2 = 1.0;
    float m2 = 0.0;
    float n = 0.0;
    for (int i = 0; i < 200; i++)
    {
        dz2 *= 9.0 * qLength2(qSquare(z));
        z = qCube(z) + kc;
        m2 = qLength2(z);
        if (m2 > 256.0) break;
        n += 1.0;
    }
    float d = 0.25 * log(m2) * sqrt(m2 / dz2);
    return float2(d, n);
}
\end{lstlisting}

\verb|calcNormal| (Listing~\ref{code:julia_normals}) implements Equation~\ref{eq-gradient} to approximate  the gradient $\nabla d$ of the distance function $d$ (Equation~\ref{eq-distance}). $\nabla d(p)$ aligns with the normal at $p$ of the isosurface with a regular value $d(p)$.

\begin{lstlisting}[caption={Normal calculation.}, label={code:julia_normals}, language=rtgCode,
	morekeywords={[1]in},
	morekeywords={[2]calcNormal, dist}]
float3 calcNormal(in float3 p)
{
    float h = 0.5773f * kPrecis;
    const float2 v = float2(1.0f, -1.0f) * h;
    return normalize(
        v.xyy * dist(p + v.xyy).x + v.yyx * dist(p + v.yyx).x +
        v.yxy * dist(p + v.yxy).x + v.xxx * dist(p + v.xxx).x );
}

\end{lstlisting}

Listing~\ref{code:closest} shows the closest-hit shader for the Julia Set. It uses a traditional approach, defining an albedo for the Phong model, combining it with a reflection color, and accumulating it with previous reflections.

\begin{lstlisting}[caption={Closest-hit shader.}, label={code:closest}, language=rtgCode,
	morekeywords={[1]in, inout},
	morekeywords={[2]shader, ClosestHit_Julia, HitWorldPosition, ObjectRayPosition, WorldRayDirection, colorSurface, reflect, WorldRayDirection, TraceRadianceRay, FresnelReflectanceSchlick, CalculatePhongLighting},
	morekeywords={[3]ProceduralPrimitiveAttributes, RayPayload, Ray}]
[shader("closesthit")]
void ClosestHit_Julia(inout RayPayload rayPayload,
    in ProceduralPrimitiveAttributes attr)
{
    // Albedo
    float3 hitPosition = HitWorldPosition();
    float3 pos = ObjectRayPosition();
    float3 dir = WorldRayDirection();
    float4 albedo = float4(3.5 * colorSurface(pos, attr.color.xy), 1);
    if (rayPayload.recursionDepth == MAX_RAY_RECURSION_DEPTH - 1)
       albedo += 1.65 * step(0.0, abs(pos.y));
    
    // Reflection
    float4 reflectedColor = float4(0, 0, 0, 0);
    float reflecCoef = 0.1;
    Ray reflectionRay = { hitPosition,
        reflect(WorldRayDirection(), attr.normal)};
    float4 reflectionColor = TraceRadianceRay(reflectionRay,
        rayPayload.recursionDepth);
    float3 fresnelR = FresnelReflectanceSchlick(WorldRayDirection(),
        attr.normal, albedo.xyz);
    reflectedColor= reflecCoef * float4(fresnelR,1) * reflectionColor;
    
    // Final color.
    float4 phongColor = CalculatePhongLighting(albedo, attr.normal);
    float4 color = phongColor + reflectedColor;
    color += rayPayload.color;
    rayPayload.color = color;
}
\end{lstlisting}

\subsection{Mandelbulb}

In Section~\ref{s-mandelbulb}, we defined the Mandelbulb fractal of the polynomial function $f(p)=p^n+c$. Remember that a point $c\in\mathbf{R}^3$ belongs to the Mandelbulb if the recurrence $f^k(0)=(f^{k-1}(0))^n+c$ does not diverges. The most popular choice in the fractal community for rendering is $n=8$.

The implementations of the Mandelbulb and Julia set (~\ref{s-julia_implementation}) are very similar. The major difference is in the function \verb|dist|, which computes the iterations using Equation~\ref{eq-white_formula} to estimate the distance function. Listing~\ref{code:mandelbulb_dist} shows the code.

Given the point $c$, the loop (lines 7-17) iterates $(0,0,0)$ using the formula $f(p)=p^8+c$. The $8$th power of $p$ is computed in lines 10-13 using the formula given in Equation~\ref{eq-white_formula}. The iterations are accumulated in line 13. If the new point gets away, we stop the loop in line 16. The distance is computed in line 19 using the formula in Equation~\ref{eq-distance} because its derivation can be applied to the Mandelbulb. Parameters used to define the colors for different parts of the fractal in the closest-hit shader latter on are also computed in lines 5 and 14.

\begin{lstlisting}[caption={Mandelbulb distance.}, label={code:mandelbulb_dist}, language=rtgCode,
	morekeywords={[1]in, out},
	morekeywords={[2]dist, pow, lenght, acos, atan2}]
float dist( in float3 c, out float4 resColor )
{
    float3 w = c;
    float m = dot(w,w);
    float4 colorParams = float4(abs(w),m);
    float dz = 1.;
    for( int i=0; i<4; i++ )
    {
        dz = 8.0*pow(sqrt(m),7.0)*dz + 1.0;
        float r = length(w);
        float b = 8.0*acos( w.y/r);
        float a = 8.0*atan2( w.x, w.z );
        w = pow(r,8) * float3(sin(b)*sin(a),cos(b),sin(b)*cos(a)) + c;
        colorParams = min( colorParams, float4(abs(w),m) );
        m = dot(w,w);
        if(m > 256.0) break;
    }
    resColor = float4(m,colorParams.yzw);
    return 0.25*log(m)*sqrt(m)/dz;
}
\end{lstlisting}

Listing~\ref{code:julia_normals} can also be used to approximate the normal vectors of the Mandelbulb surface. Basically, we just have to change the distance function by the one in Listing~\ref{code:mandelbulb_dist}.

The Mandelbulb's closest-hit shader uses the color parameters computed by the distance function and the normal. Since the code derives from several empiric tweaks of parameters weights, we avoid listing it here. Details can be found in the full shader implementation (link at the Conclusion Section).

\vspace{0.2cm}

We can create procedural scenes containing both the Julia set and the Mandelbulb. To render such a scene, we consider that each fractal is inside an AABB.
Figure~\ref{fig:example_scene} presents an example containing the Julia set, the Mandelbulb, a parallelepiped and two CSG Pac-men.

\begin{figure}[!h]
    \centering
    \includegraphics[width=0.8\textwidth]{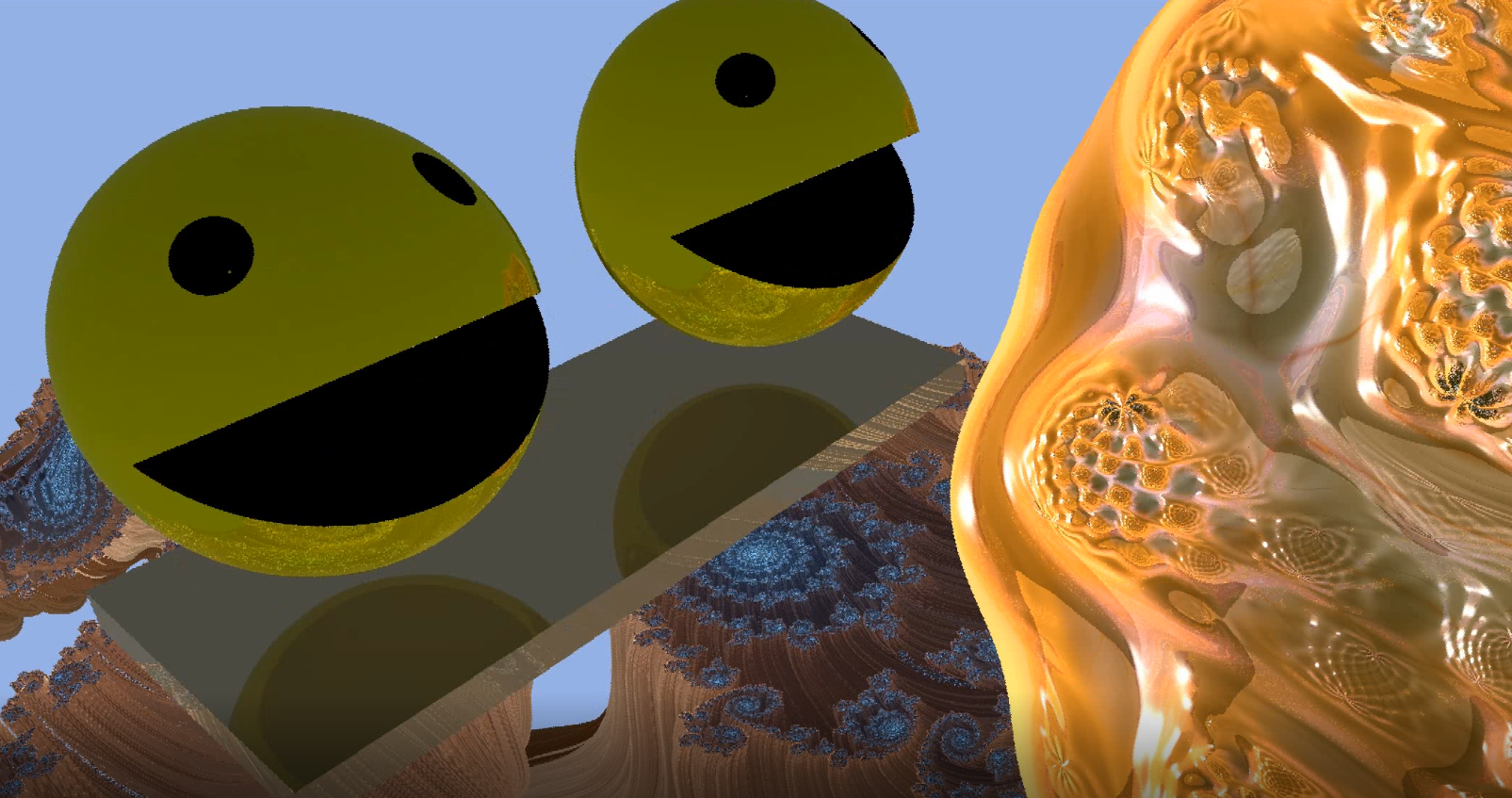}
    \vspace{-0.2cm}
    \caption{Example scene: a triangle parallelepiped mesh and several procedural objects (two Pac-men, a 3D Julia Set and a Mandelbulb).}
    \label{fig:example_scene}
\end{figure}

\section{Conclusion}
\label{sec:conclusion}

This chapter described how to use DXR shaders to implement real-time renderings of fractals based on complex dynamics. It also compiled the associated mathematical tools in an brief but intuitive way. The full code for all shaders used (including ray-generation and miss) can be found in {\footnotesize \url{github.com/dsilvavinicius/realtime\_rendering\_of\_complex\_fractals}}. For host code setup, please refer to Proceduray~\cite{dasilva2020proceduray}.

\ifdefined\BUILDRTGBOOK
\rtgBibliography{../chapter_xx/references}
\else
\rtgBibliography{fractals.bib}
\fi

\rtgEnd